\documentclass[sigconf]{acmart}

\AtBeginDocument{%
  }

\setcopyright{acmlicensed}
\copyrightyear{2018}
\acmYear{2018}
\acmDOI{XXXXXXX.XXXXXXX}
\acmConference[HPDC]{}
\acmISBN{978-1-4503-XXXX-X/2018/06}

\usepackage[font=small]{caption}
\begin{document}

\title{Parallel Data Object Creation: Towards Scalable Metadata Management in High-Performance I/O Library }



\author{Youjia Li}
\affiliation{%
  \institution{Department of Electrical and Computer Engineering, Northwestern University}
  \city{Evanston}
  \country{USA}}
\email{youjia@northwestern.edu}

\author{Rob Latham}
\affiliation{%
  \institution{Argonne National Laboratory}
  \city{Illinois}
  \country{USA}}
\email{robl@mcs.anl.gov}

\author{Robert Ross}
\affiliation{%
  \institution{Argonne National Laboratory}
  \city{Illinois}
  \country{USA}}
\email{rross@mcs.anl.gov}

\author{Ankit Agrawal}
\affiliation{%
  \institution{Department of Electrical and Computer Engineering, Northwestern University}
  \city{Evanston}
  \country{USA}}
\email{ankit-agrawal@northwestern.edu}

\author{Alok Choudhary}
\affiliation{%
  \institution{Department of Electrical and Computer Engineering, Northwestern University}
  \city{Evanston}
  \country{USA}}
\email{a-choudhary@northwestern.edu}

\author{Wei-keng Liao}
\affiliation{%
  \institution{Department of Electrical and Computer Engineering, Northwestern University}
  \city{Evanston}
  \country{USA}}
\email{wkliao@northwestern.edu}

\renewcommand{\shortauthors}{xxx et al.}

\begin{abstract}
High-level I/O libraries, such as HDF5 and PnetCDF, are commonly used by large-scale scientific applications to perform I/O tasks in parallel.
In order to provide the self-describing feature, these I/O libraries store the metadata such as data types, dimensionality, annotations, and others, along with the raw data in the same files.
While these libraries are well-optimized for concurrent access to the raw data, they are designed neither to handle a large number of data objects efficiently nor to create different data objects independently by multiple processes, as they require applications to call data object creation APIs collectively with consistent metadata among all processes.
Applications that process data gathered from remote sensors, such as particle collision experiments in high-energy physics, may generate data of different sizes or structures from different sensors and desire to store them as separate data objects.
For such applications, the I/O library's requirement on collective data object creation can become very expensive, as the cost of metadata consistency check increases with the metadata volume as well as the number of processes.
To address this limitation, using PnetCDF as an experimental platform, we investigate solutions in this paper that abide the netCDF file format, as well as propose a new file header format that enables independent data object creation.
The proposed new file header consists of two sections, an index table and a list of metadata blocks.
The index table contains the reference to the metadata blocks and each block stores metadata of objects that can be created collectively or independently.
Compared to the methods using the original header format, this new design achieves a scalable performance, cutting data object creation times by up to 582$\times$ when running on 4096 MPI processes to create 5,684,800 data objects in parallel.
Additionally, the new method reduces the memory footprints, with each process requiring an amount of memory space inversely proportional to the total number of processes.
\end{abstract}



\maketitle

\section{Introduction}
Managing large volumes of metadata for datasets stored in a file can present notable challenges with current parallel I/O libraries. Metadata refers to the information that describes the structure, organization, and properties of the raw data stored in a file. It provides essential details such as the dimensionality, data types, layout, annotation, and relationship among raw data objects, such as a hierarchical structure.
In typical use cases, metadata is relatively small in volume, and data objects are regularized with predictable dimensions and data types. However, some applications create a large number of data objects of different dimensionalities and types in a single file so it can be used for data processing tasks by other applications.
While existing parallel high-level I/O libraries have been developed to support accessing raw data efficiently, where the metadata by nature is shared among all MPI processes, they are not designed for creating a large number of unique data objects by multiple processes independently.
Such I/O demands are often seen from applications that process data collected from multiple remote sensors.
An example is the particle collider experiments in High Energy Physics (HEP) \cite{farrell2018novel, hewes2021graph, ju2021performance}.
Using a fixed-size time interval, the data collected by a sensor at a given interval consists of all the particle activity detected.
Data with the same time stamp is processed as an indivisible unit, which can be of different sizes from others.
In this case, the volume of data objects and their metadata is proportional to the number of sensors and the duration of the experiment.
Large-scale applications similar to this pose a challenge of efficiently managing and writing large volumes of metadata in parallel for modern I/O libraries.

In order to maintain the consistency of metadata among all processes when creating data objects in parallel, existing solutions adopted by the parallel I/O libraries, like PnetCDF \cite{li2003parallel} and HDF5 \cite{hdfgroupHDF5Library}, require all application processes to collectively invoke the library's API calls with identical arguments.
To fulfill this requirement, one intuitive solution is for the application programs to broadcast and synchronize metadata across all processes before calling the library's APIs.
Once the metadata of all processes are obtained by each process, its consistency check, if performed, can be expensive.
We refer to this approach as the application-level baseline approach in Section \ref{sec:app-level}.
Alternatively, Lee et al. \cite{lee2023case} approaches this problem by having each process save the data objects it creates to a unique file, i.e. the one-file-per-process approach, first and then concatenating them into one file by running a separate utility program thereafter.
This subsequent concatenation is running in sequential.
The requirement posed by PnetCDF and HDF5 on collective data object creation inherently limits the performance scalability.

This paper aims to address the aforementioned challenges by proposing novel solutions that support unique data object creation in parallel while achieving a scalable performance.
For illustration, we develop our solutions inside of PnetCDF, a high-performance parallel I/O library designed for accessing Unidata’s netCDF files \cite{ucarUnidataNetCDF}.
In this paper, we consider two kinds of data objects during their creation.
One is intended to be created by more than one process and the other by individual processes uniquely from other processes.
A parallel I/O library must be able to check the consistency of metadata among all creating processes for both kinds.
We present the following three solutions.
\begin{itemize}
\item A user application level approach that requires applications to call {\tt MPI\_Allgather} such that all processes can obtain a consistent copy of metadata before calling the library's data object creation APIs.
\item A library level approach that allows independent creation of unique data objects and checks consistency of metadata inside of I/O library. To speedup the name conflict detection, a sorting is added, which shows to be able to outperform the original hash table approach.
\item A new file header format that enables multiple MPI processes to create data objects independently and write metadata to the file header in parallel. 
\end{itemize}


To evaluate the costs of parallel data object creation, we use data generated from the Exa.TrkX project \cite{exatrkx}, a High Energy Physics (HEP) application studying the Neutrino trajectories from particle collision simulations.
In our experiments, we used a maximum of 4096 MPI processes (across 64 CPU nodes) on Perlmutter parallel computer at NERSC \cite{nerscArchitectureNERSC}.
The results presented in this paper include quantitative timing break downs to help identify the performance bottlenecks inside of the data object creation operation.
The library-level baseline approach, enhanced with a sort-based optimization for metadata consistency checks, reduces the end-to-end data object creation time by 40\% compared to the application-level baseline method.
The new file format approach with the partitioned header strategy, achieves a speedup of 582$\times$ when running 4096 processes, while other approaches did not scale up at all.

\section{Related Work}

Metadata in parallel I/O libraries like HDF5 and PnetCDF describes the structure, attributes, and organization of the raw data arrays and is stored along with the raw data in the same file.
Due to the discrepancy in HDF5 and netCDF file format, the creation of data objects is handled differently in the two libraries.

\subsection{Data Object Creation in HDF5}

HDF5 is a self-describing file format as well as an I/O library providing a set of application programming interfaces (APIs) for creating and accessing data stored in the HDF5 files. 
HDF5 also lets applications to organize data objects into a hierarchical, tree-like structure.
In HDF5, "dataset" is a term referring to the primary bulk of data in an HDF5 file, which is commonly represented in the form of a multi-dimensional array.
A dataset can have descriptive metadata in text or numerical forms to provide contextual information.
HDF5 users can also define groups, where a group is a container of one or more datasets.
Groups can also contain other groups, which is a foundation for describing a hierarchical relationship among datasets.
The HDF5 I/O library has both sequential and parallel modes.
The latter makes use of MPI-IO to support parallel data reading and writing operations to a shared file among multiple processes.


In parallel HDF5, because all data objects to be created will be stored in a single file shared among all processes, their creations are required to be collective, using the identical metadata \cite{hdfgroupHDF5Collective, hdfgroup}.
Specifically, data object creation APIs, such as {\tt H5Dcreate} (for dataset), {\tt H5Acreate} (for attribute) and {\tt H5Screate} (for dataspace), require all application processes to participate the calls and provide identical input argument values.
Recent updates \cite{soumagne2021accelerating} to the library mentions a progress to lift this constraint for the Distributed Asynchronous Object Storage (DAOS) file system, where independent data object creation can be enabled with the new API {\tt H5daos\_set\_all\_ind\_metadata\_ops}.
However, this currently seems to apply to DAOS file systems only.
It is also not documented  whether metadata consistency check is enabled or how it is implemented.

\subsection{Data Object Creation in PnetCDF}

NetCDF \cite{ucarUnidataNetCDF} is a self-describing file format and the supporting library designed for storing scientific data.
Unlike HDF5 that allows applications to describe a hierarchical relationship among data objects, netCDF follows a flat structure, meaning all variables and dimensions are stored in a single top-level namespace.
In netCDF, a raw data is referred to as "variable", which is equivalent to "dataset" in HDF5.
A netCDF variable is represented as a multi-dimensional array.
NetCDF dimensions define the shape of the arrays and can be shared by multiple variables.
Attributes in a netCDF file are associated with variables or the entire file, which is used mainly for annotations purposes. 
The classic netCDF file formats (CDF-1, 2, and 5 formats) separate metadata and raw data into distinct sections within the file: the header section stores metadata such as variable names, dimensions, and attributes, while the data section contains the raw data of the variables.
The netCDF file format is illustrated in Figure \ref{fig:file-format}. 

PnetCDF library \cite{li2003parallel} extends the capabilities of netCDF by supporting high-performance parallel I/O. Built on top of the MPI-IO standard, PnetCDF enables multiple processes to concurrently read from or write to a single netCDF file, which makes it well-suited for large-scale scientific workflows in the HPC environments. PnetCDF, like netCDF, operates in two distinct modes: define mode and data mode.
Storing data in PnetCDF involves a systematic process that begins with creating the file, which initially places the application programs in the define mode.
In this mode, users can create data objects by defining the dimensions and use them to define variables.
Once all the data objects are created, the application program transitions from define mode to data mode by calling {\tt ncmpi\_enddef}.
Inside that call, the root process writes file metadata.
We refer to this step as "end-define" stage which signifies the completion of data object creation by the application, a critical stage that can effectively check metadata consistency as to be discussed in later sections.
When creating new data objects, PnetCDF requires the API calls to be collective. Specifically, functions such as  {\tt ncmpi\_def\_dim} (for creating dimensions), {\tt ncmpi\_def\_var} (for creating variables), and {\tt ncmpi\_put\_att} (for creating attributes) require all processes to invoke the APIs using an identical argument values.
In PnetCDF, reading metadata from the file is also done collectively and occurs only when opening a file with a call to API {\tt ncmpi\_open}.
During that API call, the entire file header is retrieved from the file and stored in an internal buffer replicated among all processes.

\section{Design and Implementation}
\label{sec:design}

To address the parallel data object creation challenge, we first consider a method that simply exchanges and synchronizes metadata at the application side before invoking the parallel I/O data object creation APIs.
This application-level baseline approach serves as a base method for comparison.
The second method moves the metadata synchronization into the I/O library, alleviating applications from the burden of performing the synchronization. 
This method also enables independent object creation.
In other words, it allows different data objects to be created by different processes in parallel.
The library-level baseline approach must ensure metadata consistency for the same data objects to be created by two or more processes. 
When the number of data objects or the number of processes becomes large, the cost of metadata synchronization can apparently be expensive.
At last, to reduce such cost, we introduce a new file header format, which largely removes the need for metadata synchronization and allows all processes to write to the disjoint header sections concurrently. 

In the rest of this paper, the terms "define" and "create' may be used alternatively while referring to creating a data object and its metadata.
Below is a list of the technical terminology to be used in the rest of the paper.
\begin{description}
\item[Data object] is a list of numerical values storing the measurements of a phenomena collected from an experiment. They are often organized into a multi-dimensional array. Examples are temperatures measured within a geographical region over a certain period of time. We consider two kinds of data objects when creating them in parallel.
\item[Metadata] is the data describing the data objects. Examples are dimensionality, data type, and annotation. The volume of metadata is usually much less than the data objects they describe.
\item[Shared data objects] are the data objects to be created by two or more processes. They are intended to be accessed later by multiple processes. In both PnetCDF and parallel HDF5, all data objects (namely variables in PnetCDF and datasets in HDF5) are shared and required to have consistent metadata when created in parallel.
\item[Non-shared data objects] are the data objects to be uniquely created by a single process. They are intended to be accessed later only by the creating processes only.
Once non-shared data objects have been created and their metadata written in the file, they can be accessed by the non-creating processes through inquiring their IDs first.
\item[Consistency] is the state of metadata of data objects when created in parallel. The shared data object's metadata should be the same among the calling processes. On the other hand, the metadata of non-shared data objects should be unique by themselves. For non-shared data objects, the I/O libraries must check for name conflict. The data objects with the same name are identified as shared objects and the I/O libraries must check their metadata contents for consistency.
\item[End-define] is the operation phase for applications to indicate all the new data objects to be created in the file have completed. This is implemented explicitly in PnetCDF through its API {\tt ncmpi\_enddef}, which must be called collectively. At the end of this API call, all the metadata is written to the file. HDF5 does not implement such phase. 
\end{description}


\subsection{Application-level Baseline Approach}
\label{sec:app-level}

\begin{figure}[t]
  \centering
  \includegraphics[width=\linewidth]{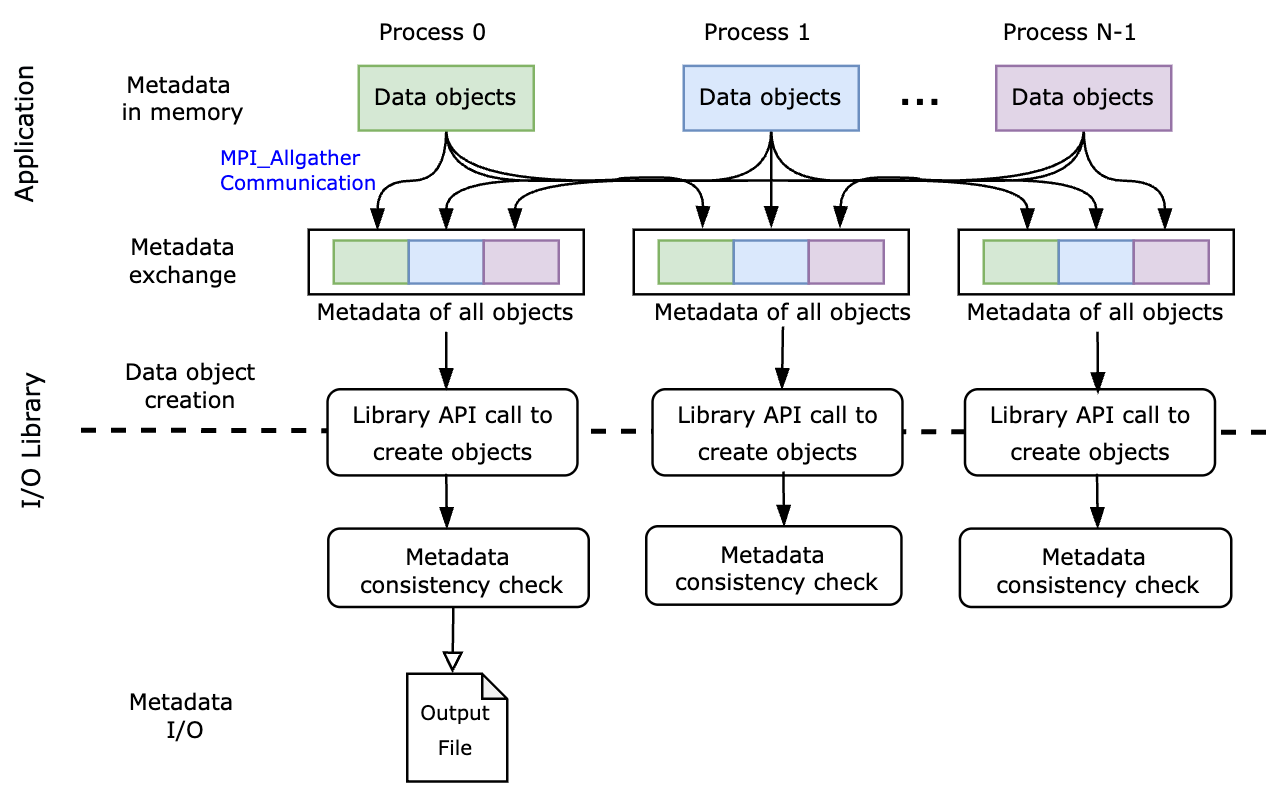}
  \caption{Metadata flow in the application-level baseline approach. Abiding by the I/O library's requirements that data objects must be collectively created by all processes using a consistent copy of metadata, user applications explicitly exchange and synchronize metadata first before making the I/O library API calls.}
  \label{fig:app-level}
\end{figure}

To fulfill the requirements of parallel I/O libraries on collective data object creation with consistent metadata, one intuitive approach is the applications explicitly make MPI communication calls to synchronize the metadata, if the metadata is different among processes.
A schematic for metadata flow among processes is shown in Figure \ref{fig:app-level}.
An application program launches an {\tt MPI\_All\_gather} call to synchronize metadata before invoking the library APIs.
To minimize the number of calls to {\tt MPI\_All\_gather}, we first serialize the metadata into a single send buffer.
The data object creation process of this baseline approach consists of the following phases.:
\begin{itemize}
\item Phase 0 - Metadata synchronization: A call to {\tt MPI\_All\_gather} to gather the size of local metadata of all processes, so a buffer to be used to receive the metadata can be allocated. Then, another call to {\tt MPI\_All\_gather} to gather all the metadata in each process. Metadata consistency is checked on each process and returns an error when an inconsistency is found.
\item Phase 1 - Data object creation: the application program calls the I/O library's API to create data objects using the consistent metadata.
\item Phase 2 - End-define: the application program calls the I/O library API to exit the define mode, if such API is available. Note the I/O library may perform its own metadata consistency check if it is implemented in the library. For instance, PnetCDF performs such check in its end-define stage if users enable the safe mode. At the end the root process writes the metadata to the header section of the file.
\end{itemize}

\subsection{Library-level Baseline Approach}
\label{sec:lib-level}

\begin{figure}[t]
  \centering
  \includegraphics[width=\linewidth]{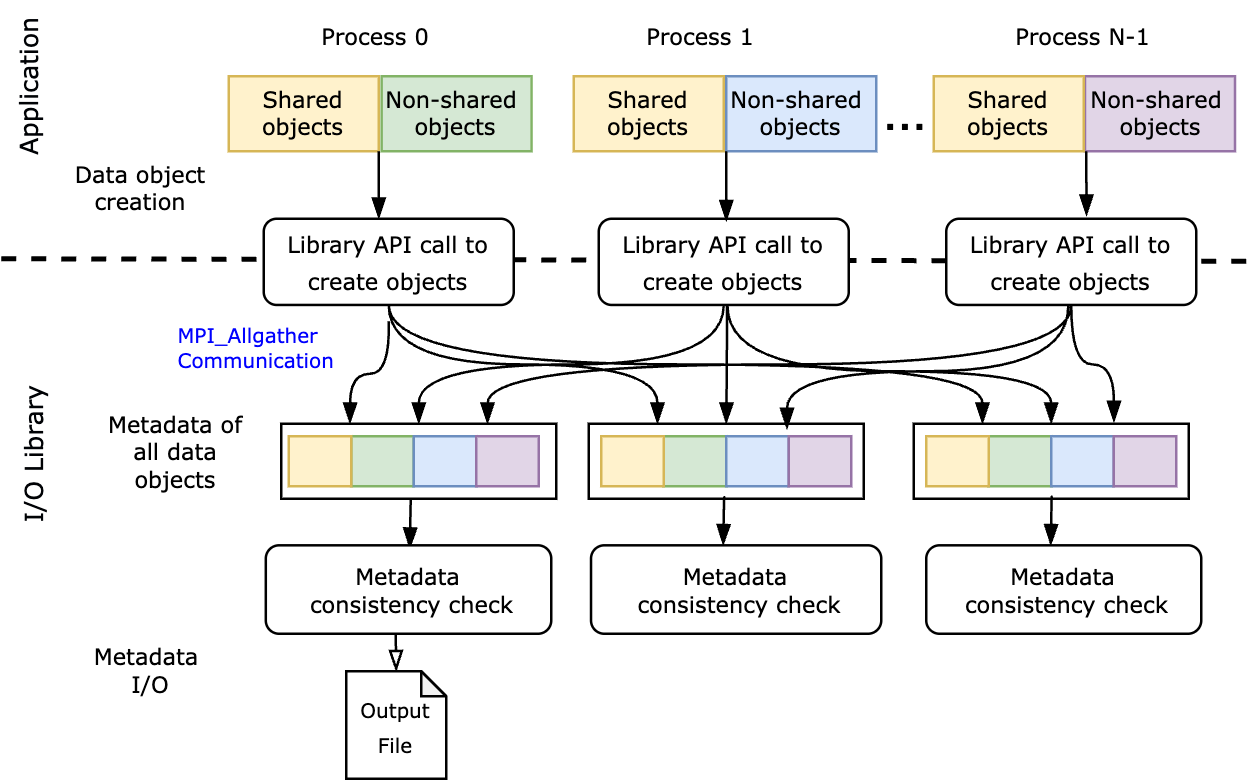}
  \caption{Metadata flow in the library-level baseline approach. I/O library is modified to allow shared and non-shared objects to be created collectively or independently. Applications reply on the I/O library to check the metadata consistency.}
  \label{fig:lib-level}
\end{figure}
To relieve the burden of explicitly performing metadata synchronization from the users and allow processes to create unique data objects from others, we propose the library-level baseline approach.
This approach still supports shared data objects created collectively by multiple processes. In addition, it allows the creation of non-shared objects independently by individual processes.
The library is responsible to automatically detect shared and non-shared data objects and synchronize metadata for the shared objects.

Compared to the application-level baseline approach, this approach handles metadata synchronization within the I/O library.
It eliminates the constraints of collective API invocations for data object creation. 
The fact that processes can independently create new objects with different metadata provides application developers a flexibility. 
A schematic for metadata flow under this approach is shown in Figure \ref{fig:lib-level}. 
The I/O library merges metadata from all processes and check their consistency at the end-define stage, after which the root process performs the actual I/O by writing the entire metadata into the file header. 
As the metadata consistency check can potentially become one of the major performance bottlenecks when the number of data objects is large, we also develop a sort-based improvement for name checking, to be discussed in Section \ref{sec:consistency}. 
The data object creation process under this approach skips Phase 0 (metadata synchronization in the application-level baseline approach) and consists of the following phases:

\begin{itemize}
\item Phase 1 - Data object creation: each process calls the library APIs to create all data objects. No explicit communication to synchronize metadata is necessary.
\item Phase 2 - End-define: with all the data objects and their metadata known, this phase is decomposed into:
\item Phase 2-0  Metadata synchronization: similar to Phase 0 of the application-level baseline approach, it makes MPI communication calls to make metadata available on all processes.
\item Phase 2-1 Metadata consistency check: each process first identifies the shared data objects if their names are the same. For shared data objects, it further checks the consistency of metadata contents. An error is returned if an inconsistency is found.
\item Phase 2-2 Metadata I/O: the consistent metadata is written to the file by the root process.
\end{itemize}


\subsection{Metadata Consistency Check}
\label{sec:consistency}

Metadata consistency refers to ensuring that every data object is uniquely created.
The unique identifier of a data object is its name, including the path if a hierarchical representation is used.
Currently, both PnetCDF and HDF5 require data object creation to be collectively, using the identical metadata among all processes.
In the proposed library-level baseline approach, we relax this requirement to allow individual processes to create new data objects independently from other processes if the object's name (and its path) is unique.
A data object can also be created collectively by multiple processes as long as its metadata is consistent among the calling processes.
A parallel I/O library may choose to implement a consistency check mechanism to detect any inconsistent metadata for all data objects created in parallel.
Based on our understanding, the parallel HDF5 library currently (version 1.14.5) does not perform metadata consistency check.
In cases where divergences occur, the metadata created by the root process is the authoritative version written to the file.
The PnetCDF library performs metadata consistency check only when the "safe mode" is enabled by the user.
In both cases, as all the data object creation APIs are required to be collective and using the same argument values, the consistency is checked at each API call by having the root process broadcasting its metadata, followed by all processes comparing it against their own copy of metadata.
An error code is returned if an inconsistency is found.
This mechanism ensures that metadata with the same name but different in metadata are detected as an error.
Although such checking helps user applications to identify the intended or unintended data consistency, it can become expensive when the number of data objects or application processes is large.

To support independent data object creation, our proposed library-level baseline approach must implement a mechanism to handle situations where objects with the same name have different metadata.
The metadata consistency check is deferred to the ``end-define'' stage when all the data objects to be created are known, so the library can start checking any conflict.
At first, the library examines all object names gathered from all processes.
If data objects with identical names are found, they are considered as shared objects and their metadata are compared.
If the metadata differ, the library raises a name inconsistency error.

Metadata consistency checks may come with non-negligible runtime overheads, particularly for large amount of metadata or large-scale runs. To check name conflicts, the PnetCDF library uses a hash table for name quick lookup.
While this method is efficient for small to moderate metadata volumes, its performance can degrade significantly for large numbers of objects when the hash table size is not big enough causing too many hash key collisions.
An ideal hash function should minimize the hashing collision, which is often achieved by tuning the hash table size. 

When iteratively inserting $n$ objects into a hash table of size $k$, each slot on average stores $n/k$ objects assuming a uniform hash distribution. The first insertion to the slot requires zero string comparison cost, and the last one makes $(n/k - 1)$ comparisons, so the average cost per insertion is the arithmetic mean, resulting in approximately $n/2k$ string comparisons per insertion.
The time complexity for name string comparisons is given by the following formula.
\begin{equation}
 O(n) = n \cdot \frac{n}{2k}
 \label{eqn:str_compare}
\end{equation}

While increasing the hash table size or implementing a rehashing mechanism can reduce the likelihood of hash collisions, the runtime is still cursed by the potentially large number of checks upon every insertion on new object names, as reflected in the first $n$ factor in Equation (\ref{eqn:str_compare}).
We propose a sorting-based optimization to eliminate the dependency on hash tables. 
With all the metadata known at the end-define stage, it provides an opportunity to perform a one-time scan for the entire list of new objects. Building on this insight, we sort the data object names into an increasing order.
Given a sorted list, name conflicts can be efficiently detected by checking every two adjacent elements.

\subsection{Data Object Identifier Mapping}
\label{sec:id-map}

\begin{figure}[t]
  \centering
  \includegraphics[width=\linewidth]{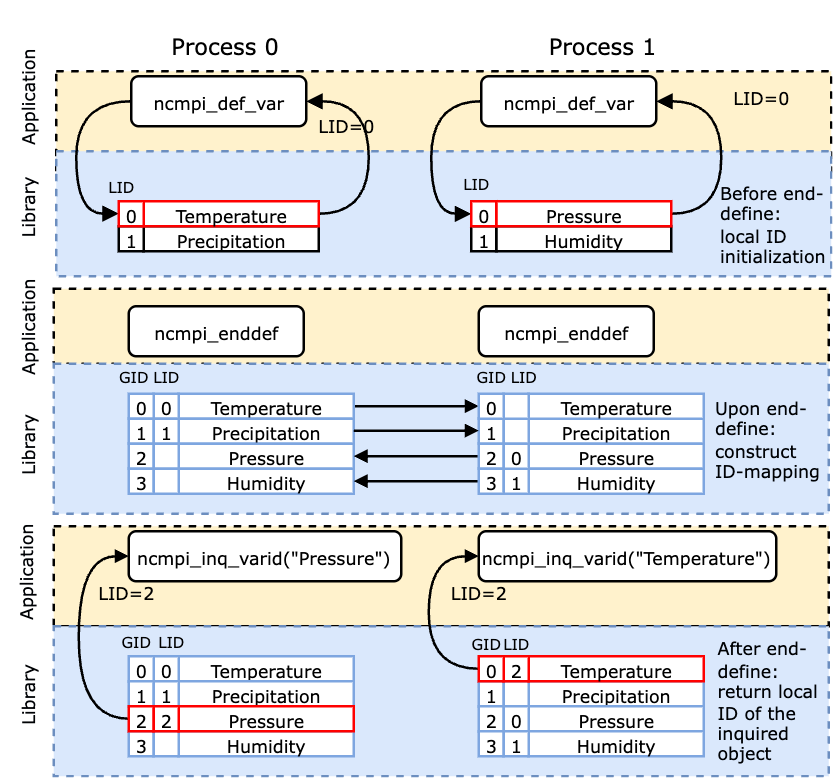}
  \caption{Data object identifier mapping mechanism. LID (Local ID) is the ID returned by the I/O library to an application process. GID (Global ID) is statistically mapped to the data objects stored in the file and used internally by the library.}
  \label{fig:id-map}
\end{figure}

The HDF5 library supports a data object ID mapping mechanism in the back-end to allow each process to hold a different integer ID for the same object.
In PnetCDF on the other hand, the IDs of the same data object are kept identical among all processes.
An object ID is simply the index pointing to the array storing objects of the same kind (e.g. dimensions or variables), based on the order of their creation.
However, in the context of independent object creation, it is less straightforward to determine the object ID returned to the user because the global index is unknown to each process until metadata exchange at "end-define" stage.
For applications, the object ID obtained from PnetCDF prior to "end-define" stage needs to remain valid for future references.
To tackle this issue, we design the following global-local ID mapping mechanism:
\begin{itemize}
\item Global IDs are the IDs of data object that have been created in the file. Global IDs must be identical among all processes. 
\item Local IDs are the IDs returned by the I/O library. For each process, newly created data objects are always assigned to a locally unique ID.
Once assigned, local IDs are persistent until the file is closed. Local IDs are not portable across processes.
\end{itemize}

Figure \ref{fig:id-map} illustrates a simple example of ID mapping before and after the "end-define" stage.
Initially when defining variables temperature and pressure, processes 0 and 1 are returned the local IDs of both 0 from the I/O library.
After the "end-define" stage, both processes 0 and 1 obtain a local ID of value 2 when inquiring/opening the variables pressure and temperature respectively created by the other process earlier.
Global IDs are used by the I/O library internally only and not visible to the user applications.
They are used to uniquely identify the data objects stored in the file.
The conversion of local and global IDs is done implicitly inside the library.

\subsection{New Header Format Approach}
\label{sec:new-format}

\begin{figure}[t]
  \centering
  \includegraphics[width=\linewidth]{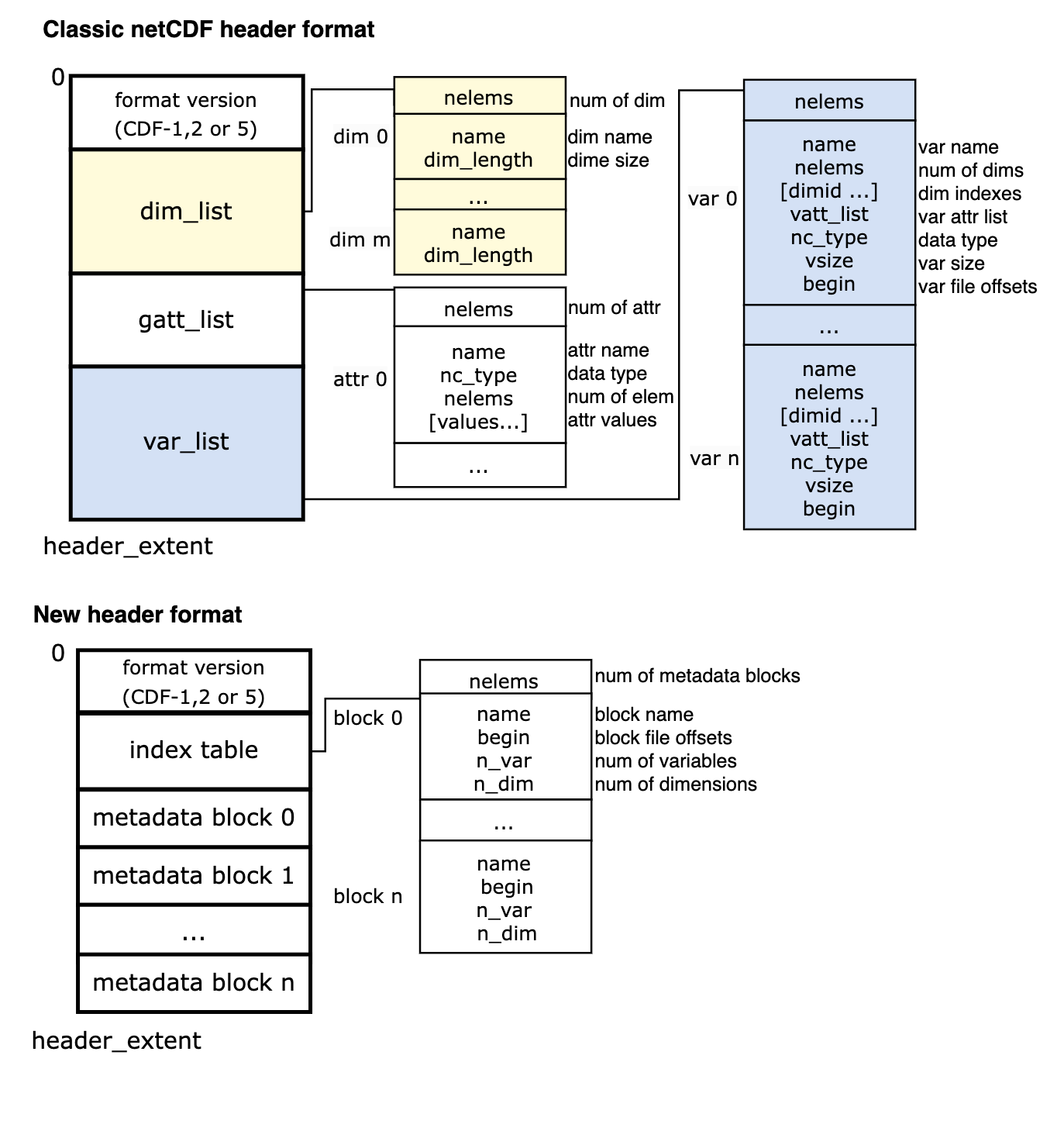}
  \caption{Top figure shows the original classic netCDF file header format. Bottom shows the proposed new header format, which consists of an index table and a list of metadata blocks. Metadata blocks essentially follow the original netCDF header format.}
  \label{fig:file-format}
\end{figure}

\begin{figure*}[t]
  \centering
  \includegraphics[width=\linewidth]{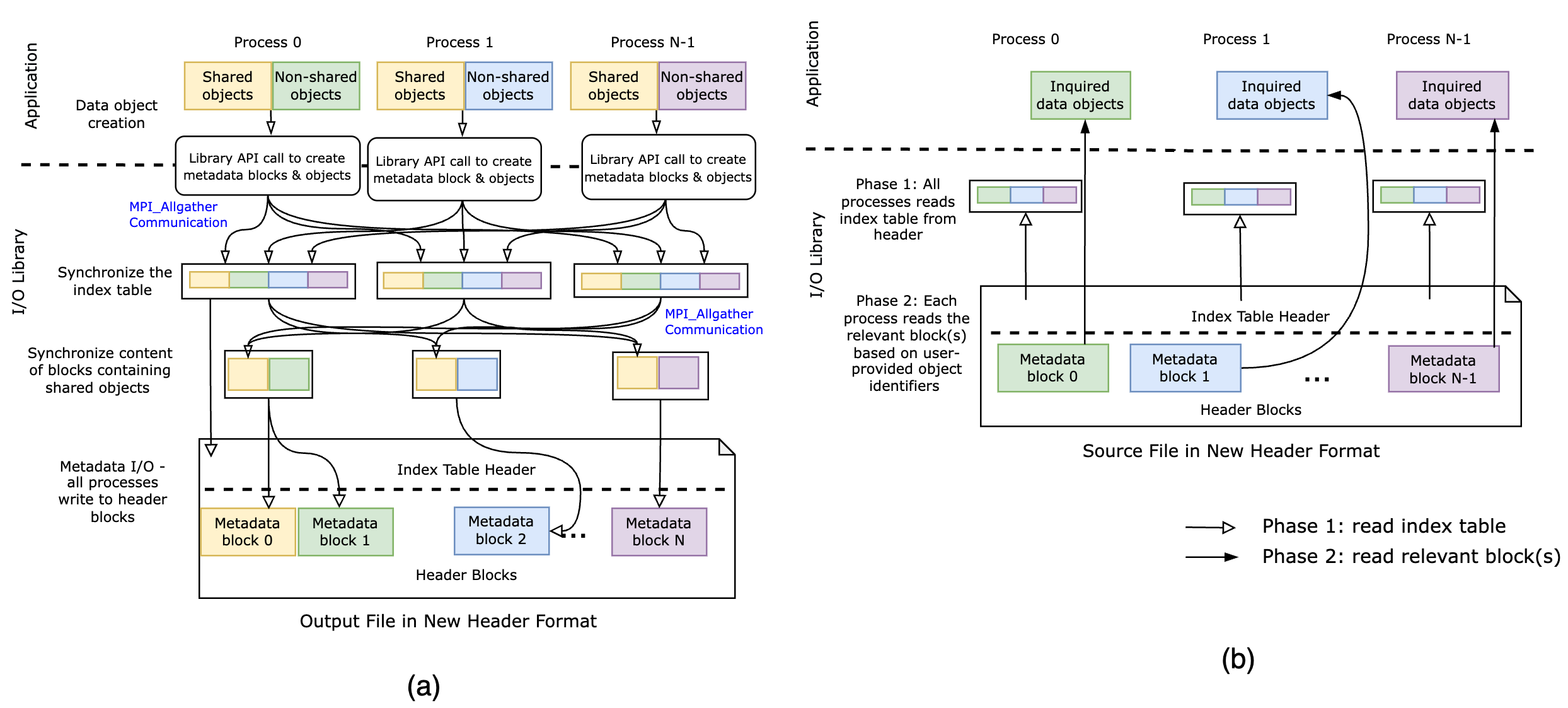}
  \caption{Metadata flow for new header format approach. (a) During data object creation, metadata is synchronized among processes for both shared and non-shared objects. Metadata consistency check is performed in two steps. At first, data object names are checked for conflicts. Secondly, for data objects with the same name, identified as shared objects, their metadata are checked for consistency. (b) For read operations, only those metadata blocks containing the requested data objects are retrieved from file.}
  \label{fig:new-header-flow}
\end{figure*}

To support independent data object creation, all the approaches discussed by far require a copy of entire metadata to be made available on all processes for consistency check and only the root process writes the metadata to the file.
Apparently, such strategy is not scalable for large applications.
In case when a data object uniquely created by a process, making the metadata of such data objects available on all processes for consistency check is not necessary.
We propose a new file header format that can avoid such unnecessary metadata consistency check and allow all processes to write metadata to the file header in parallel.
The new format consists of two sections, an index table and a list of metadata blocks.
The index table stores the information about individual metadata blocks and each metadata block contains the metadata of data objects created by one or more application processes.
A metadata block is uniquely identified by its name, which is a string path provided by applications when calling the library APIs to create a data object.
The idea of using metadata blocks thus simplifies the consistency check to those blocks created by two or more processes.

\begin{table*}[t]
  \caption{Statistics of the two data sets used in the performance evaluation}
  \label{tab:dataset}
  \small
\begin{tabular}{|lccccccc|} \hline
\multicolumn{8}{|c|}{\textit{dataset\_98M}} \\ \hline
\multicolumn{1}{|l|}{No. of processes}                              & \multicolumn{1}{c|}{Total}   & \multicolumn{1}{c|}{4}       & \multicolumn{1}{c|}{16}     & \multicolumn{1}{c|}{64}     & \multicolumn{1}{c|}{256}   & \multicolumn{1}{c|}{1024} & 4096 \\ \hline
\multicolumn{1}{|l|}{No. of netCDF Variables assigned per process}  & \multicolumn{1}{c|}{568480}  & \multicolumn{1}{c|}{142120}  & \multicolumn{1}{c|}{35530}  & \multicolumn{1}{c|}{8883}   & \multicolumn{1}{c|}{2221}  & \multicolumn{1}{c|}{555}  & 139  \\ \hline
\multicolumn{1}{|l|}{No. of netCDF Dimensions assigned per process} & \multicolumn{1}{c|}{852715}  & \multicolumn{1}{c|}{213179}  & \multicolumn{1}{c|}{53295}  & \multicolumn{1}{c|}{13324}  & \multicolumn{1}{c|}{3331}  & \multicolumn{1}{c|}{833}  & 208  \\ \hline
\multicolumn{1}{|l|}{Metadata amount assigned (MB)}                  & \multicolumn{1}{c|}{70.72}   & \multicolumn{1}{c|}{17.68}   & \multicolumn{1}{c|}{4.42}   & \multicolumn{1}{c|}{1.11}   & \multicolumn{1}{c|}{0.28}  & \multicolumn{1}{c|}{0.07} & 0.02 \\ \hline
\multicolumn{8}{c}{} \\ \hline
\multicolumn{8}{|c|}{\textit{dataset\_1G}} \\ \hline
\multicolumn{1}{|l|}{No. of processes}                              & \multicolumn{1}{c|}{Total}   & \multicolumn{1}{c|}{4}       & \multicolumn{1}{c|}{16}     & \multicolumn{1}{c|}{64}     & \multicolumn{1}{c|}{256}   & \multicolumn{1}{c|}{1024} & 4096 \\ \hline
\multicolumn{1}{|l|}{No. of netCDF Variables assigned per process}  & \multicolumn{1}{c|}{5684800} & \multicolumn{1}{c|}{1421200} & \multicolumn{1}{c|}{355300} & \multicolumn{1}{c|}{88825}  & \multicolumn{1}{c|}{22206} & \multicolumn{1}{c|}{5552} & 1388 \\ \hline
\multicolumn{1}{|l|}{No. of netCDF Dimensions assigned per process} & \multicolumn{1}{c|}{8527150} & \multicolumn{1}{c|}{2131788} & \multicolumn{1}{c|}{532947} & \multicolumn{1}{c|}{133237} & \multicolumn{1}{c|}{33309} & \multicolumn{1}{c|}{8327} & 2082 \\ \hline
\multicolumn{1}{|l|}{Metadata amount assigned (MB)}                  & \multicolumn{1}{c|}{802.20}  & \multicolumn{1}{c|}{200.55}  & \multicolumn{1}{c|}{50.14}  & \multicolumn{1}{c|}{12.53}  & \multicolumn{1}{c|}{3.13}  & \multicolumn{1}{c|}{0.78} & 0.20 \\ \hline
\end{tabular}
\end{table*}

Figure \ref{fig:file-format} illustrates the proposed new file header format.
The netCDF classic file format is provided as a comparison. 
In the new header format, metadata is partitioned into blocks so that they can be independently created and written to file by application processes. 
In netCDF, metadata of a data object, also referred to as variable, contains name, dimensions, types, and attributes \cite{netcdf_format}.
The classic netCDF file header format organizes the metadata into three lists, one for each of three metadata kinds, dimensions, attributes, and variables. 
Metadata of the same kinds are stored contiguously in its list.
Such design makes the parallel I/O to the file header difficult to implement efficiently.

In the proposed new header format, the index table stores references to all the metadata blocks.
When creating a new data object, applications can add a path name prefix to the data object's name.
This path prefix string serves the user intent for creating unique or shared data objects.
Data objects sharing the same path name are assigned to a metadata block and each block is identified by the path name.
If a data object is intended to be created uniquely by only one application process, then its name string, including the path, must be unique among all processes.
On the other hand, if a data object is intended to be created by multiple processes, then its name string and metadata contents must be identical among the creating processes.
The index table stores high-level information about all metadata blocks, including their file offsets, size, and statistics, such as the number of data objects in a block.
The index table is synchronized and replicated on all processes after the end-define stage, so that each process can independently search, identify, and locate the metadata blocks for data objects requested by the user.
The size of the index table is expected to be small, compared to the metadata blocks.
The second half section of the new file header stores a list of metadata blocks.
The format of each metadata block essentially follows the original netCDF header format.
For read operations, because a copy of the index table is cached locally on all processes, each process can selectively read the relevant metadata blocks.

The index table can also support high-level metadata inquiry.
An example is the neural network training applications which need to obtain the number of samples stored in an input file, so to determine the training batch size prior to the model training. 
If the input file is in HDF5 format, obtaining the total number of data objects can be expensive.
This is because metadata in an HDF5 file is stored in blocks of fix-sized units and metadata blocks can disperse all over the file space.
Obtaining the total number of data objects requires to retrieve all metadata blocks, resulting in many file seeking and reading operations.
With our proposed new file format, such metadata request can be served more efficiently by scanning only the index table, which is stored in a contiguous space in the file.


When implementing the new header format in PnetCDF, the following modifications are required.
At first, the library must uniquely identify each data object using a combination of the metadata block name and the data object name in its block.
APIs for creating or accessing data objects must allow users to provide a name string that includes a path.
The library uses path names to identify the shared and non-shared metadata blocks and leverages such information to perform consistency check and header I/O.
Figure \ref{fig:new-header-flow} depicts how metadata is processed in the I/O library.
At the end-define stage, the first step is to construct the index table, which collects all metadata block names from all processes.
A string comparison is then performed to tell shared metadata blocks from non-shared ones.
Metadata consistency check is only applied on data objects defined in the shared metadata blocks, which involves MPI communication to make the block contents available on the sharing processes.
Note that data object name conflicts are checked at the object creation time before the end-define stage at each process. 
The time complexity of string comparisons for the metadata blocks is presented in Equation \ref{eqn:new-format-str-compare}.
It is based on the adoption of the hash table for storing names and their lookups.
To simplify the representation without losing generality, we assume each process creates $n/p$ number of data objects, where $n$ is the total number of data objects, $p$ is the total number of application processes, and $k$ is the hash table size. Since each process inserts $n/p$ objects into its hash table, and each slot holds $n/kp$ objects on average, the insertion cost follows a linear progression from 0 to $(n/kp - 1)$ . By computing the arithmetic mean, the average comparison cost per insertion is approximately $n/2kp$. 

\begin{equation}
O(n) = O\left(\frac{n}{p} \cdot \frac{n}{2kp}\right) + O\left(p \cdot \frac{p}{2k}\right)
\label{eqn:new-format-str-compare}
\end{equation}
When $p \ll n$ (a common case), the runtime for consistency checks is significantly reduced.
However, in extreme scenarios of $p = 1 $ (a single-process program) or $p = n$ (where each process creates one data object), the time complexity becomes similar to Equation (\ref{eqn:str_compare}).

Once the consistency check completes, the sizes of metadata blocks are calculated to generate the corresponding file offsets for each block, so they can be stored in the index table.
When writing the file header, the index table is written by the root process and each metadata block is written by only one of its creator processes.
Figure \ref{fig:new-header-flow}(b) depicts the case when opening a file using the new header format.
Contrasting to the current PnetCDF implementation that reads the entire file header and duplicated among all processes, the new format approach reads and duplicates only the index table at the file open, and metadata blocks are read from file only when an application process makes requests to the data objects they contain.
Thus, retrieving the index table can be implemented using an MPI collective read operation, and reading metadata blocks can use independent read operations.

\section{Performance Evaluation}
\label{sec:eval}

To evaluate the runtime performance of the proposed approach, we perform strong-scaling experiments with intensive metadata management workloads. 
All timings reported is the maximum among processes. 
Our performance evaluations are conducted on Perlmutter \cite{nerscArchitectureNERSC}, a HPE Cray EX supercomputer at NERSC, with CPU nodes equipped with 512 GB of memory per node. 
Our experiments utilized up to 4096 MPI processes distributed on 64 nodes. 
The I/O operations are served by the Lustre parallel file system, which consists of 248 I/O servers, known as OSTs. 
In our experiments, Lustre settings have an overall small impact on metadata writing because the actual I/O amount is comparatively low. 
We configure the file striping size to 2 MiB and striping count to 8 and 64. 
We select a striping count of 8 for baseline approaches since there is only 1 writer (root process) for metadata writing. 
For the new header format approach, since all processes make I/O requests to write header, we tune up the stripe count to 64 to allow all aggregators in collective MPI-IO to efficiently write to the header section.

For performance evaluation, we utilize a data set generated by a Neutrino particle collision simulation from the Exa.TrkX project.
The sub-atomic particles produced from collisions are detected by sensors which divide the collected data by a fix-sized time interval, called event.
Data in an event contains particles of different kinds, their coordinates, mass, velocities, and many other measurements.
As the degrees of collision activities vary from time to time, the data volume of an event is different from all others.
One of the Exa.TrkX tasks is to reconstruct the particle trajectories using Graph Neural Network (GNN).
To generate input data to train GNN models, the simulation data is processed into graph samples and each sample is contains all the data of a given event.
All the graph samples contain the same kinds of data but of different sizes.
In fact, training samples of different sizes are commonly seen in GNN training in many research fields, as the graph data inherently involves dynamic and unpredictable structures \cite{yu2023towards,  zheng2022instant, rossi2020temporal}. 
Graph data are capable of modeling various real-world scenarios, including social networks \cite{kumar2019predicting, song2019session, alvarez2021evolutionary}, traffic networks \cite{yu2017spatio, wu2019graph, guo2019attention}, physics \cite{farrell2018novel,huang2020learning,pfaff2020learning} and other scientific data \cite{choudhary2021atomistic, li2024hybrid, coley2019graph}. 

As each simulation run of Exa.TrkX employs thousands of sensors running a long period of time, the data volume can be very large and thus the task of graph sample generation is parallelized to run on HPC computers.
This task first partitions the data among all processes into equal workloads, so that the graphs can be generated independently and concurrently.
The GNN model training task in Exa.TrkX project workflow takes a single input file that stores all the graph samples.
This design choice is to allow the GNN training program to construct mini-batches more efficiently than storing samples in multiple files. 
When using HDF5 or PnetCDF to create a single training file in parallel, the graph generation task must synchronize metadata of all graphs across all processes.

The simulation data set from the Exa.TrkX simulation contains 568,480 data objects.
In order to study performance of the solutions proposed in this paper for different metadata sizes, we artificially increase the data set into 10 times as the second data set.
The original data set contains metadata of a moderate size of 98 MB (labeled as $dataset\_98M$), and metadata of the augmented data set is of size 1071 MB (labeled as $dataset\_1G$). 
We evenly partition the data objects among all the processes allocated in each run, closely mimicking the scenario of the graph generation task. 
Table~\ref{tab:dataset} shows some statistics of the two data sets.

\begin{figure}[t]
  \centering
  \includegraphics[width=\linewidth]{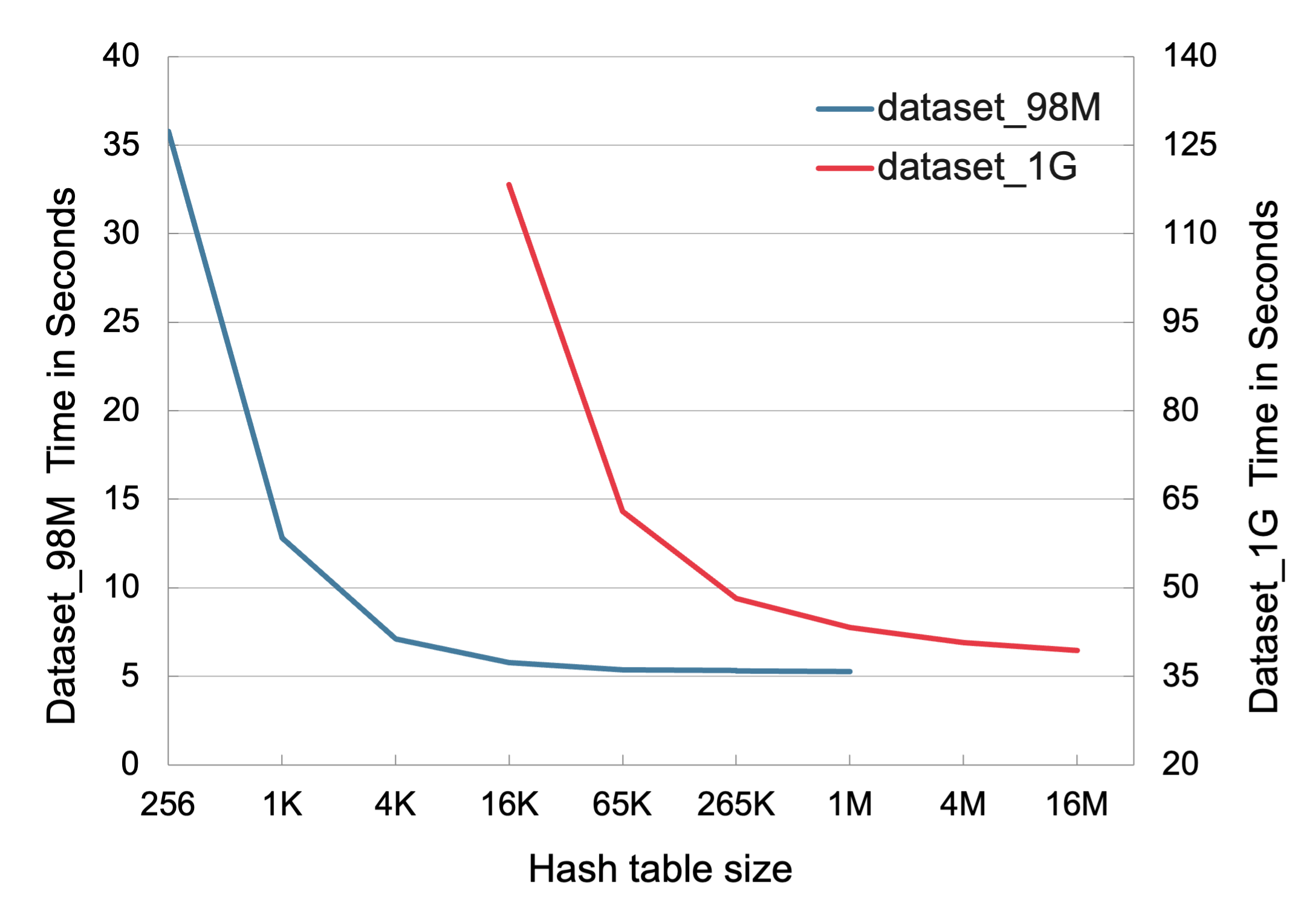}
  \caption{Data object creation times under different hash table sizes for the application-level baseline approach when running 64 processes on one compute node.}
\label{fig:hash-size}
\end{figure}

As mentioned in Section \ref{sec:consistency}, the hash table size can be an important factor affecting the data object creation time. 
At first, we would like to determine a proper hash table size that can yield a good performance, given the sizes of two data sets used in our experiments.
Figure \ref{fig:hash-size} shows the timing results of the application-level baseline approach for various hash table sizes. 
As the hash table size increases, the data object creation time reduces.
The timing curves flatten at around 16384 and 1048576 for $dataset\_98M$ and $dataset\_1G$, respectively. 
Based on these results, we set the hash table sizes to these values in all the experiments presented in this study.

\begin{figure*}[t]
  \centering
  \includegraphics[width=\linewidth]{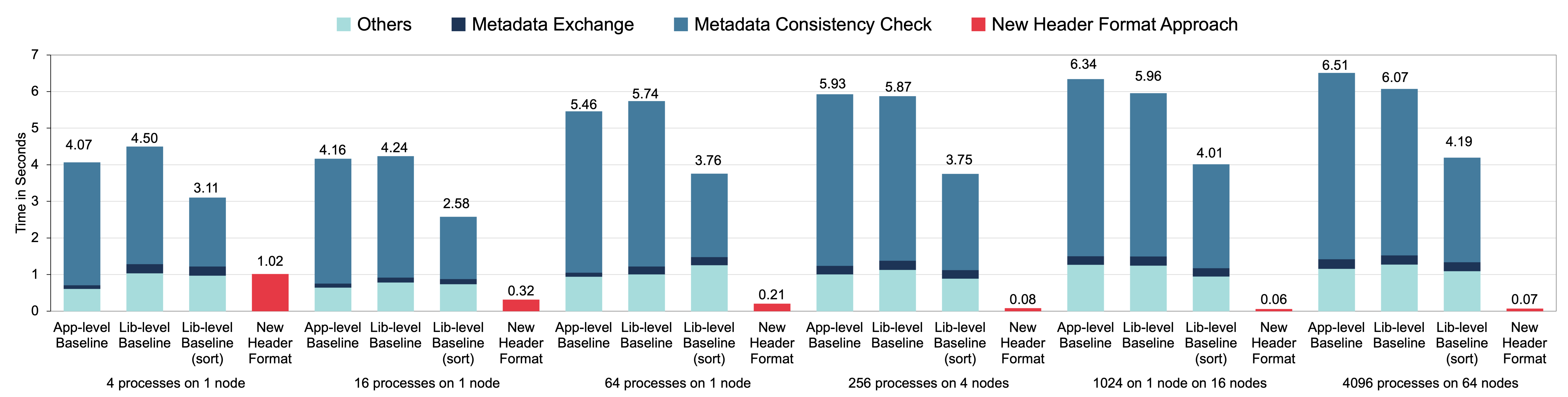}
  \caption{Timing breakdowns when using $dataset\_98M$. The timing breakdown for the new header format approach is shown in Figure \protect\ref{fig:para-breakdown1}.}
  \label{fig:write-time1}
\end{figure*}

\begin{figure*}[t]
  \centering
  \includegraphics[width=\linewidth]{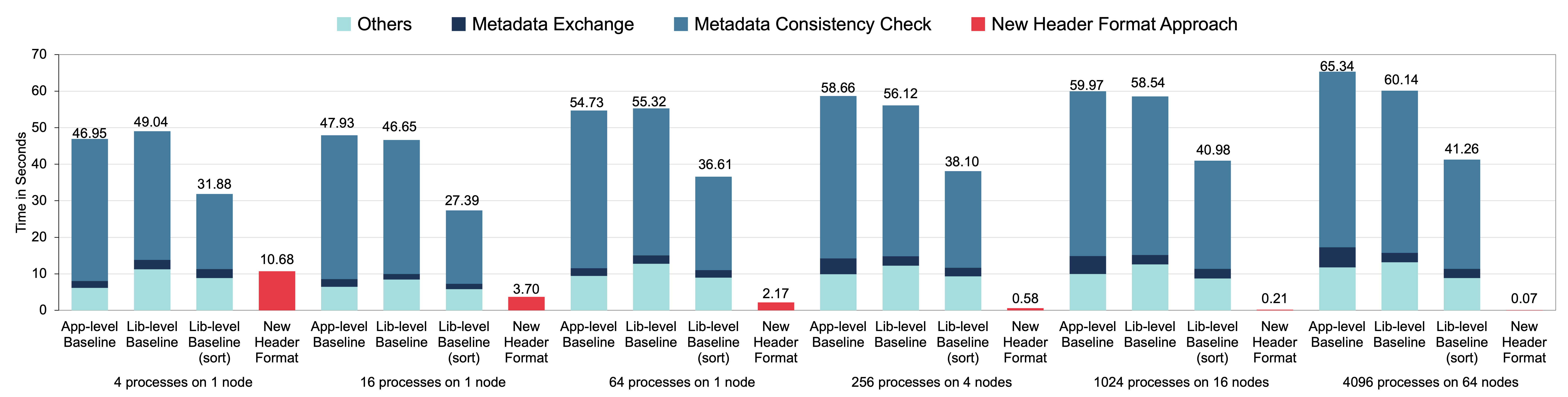}
  \caption{Timing breakdowns when using $dataset\_1G$. The timing breakdown for the new header format approach is shown in Figure \protect\ref{fig:para-breakdown2}.}
  \label{fig:write-time2}
\end{figure*}

\subsection{Baseline Approaches Performance}

In Section \ref{sec:design}, we propose 3 baseline approaches: the application-level baseline, the library-level baseline, and a variant of the library-level baseline that incorporates a sorting optimization. 
%
Figures \ref{fig:write-time1} and \ref{fig:write-time2} present the times measured from the first data object creation till the end of end-define stage.
The timing breakdowns are also included for studying the performance in details.
The timing costs are organized in to six groups, each corresponding to a run on a different number of MPI processes.
The left-most three bars in each group represent the three baseline approaches. 
We observe that all 3 baseline approaches do not scale at all, as the number of processes increases.
To help understanding the performance bottlenecks, the timing breakdowns shown in Figures \ref{fig:write-time1} and \ref{fig:write-time2} reveal that MPI communication cost (represented by metadata exchange) accounts for a small fraction of the total time for all the baseline approaches. 
Instead, the metadata consistency check dominates the end-to-end time. 
The consistency check performs pair-wise metadata comparisons for all the data objects on each process and thus its cost does not reduce as the number of process increases.
In fact, the cost is slightly increases, which may be owing to memory resource allocation contention on individual compute nodes when the number of processes running on a node is large.
When comparing the application-level and library-level (without the sorting optimization) approaches, the costs of metadata consistency check are essentially the same.
This result is expected because the same consistency check is carried out in both approaches, but just in different places.


Figures \ref{fig:write-time1} and \ref{fig:write-time2} show that
the sorting-based consistency check achieved a runtime reduction of approximately 40\% over the one without the sorting optimization. 
The improvement demonstrates that using a sorted array is more efficient for checking name conflicts when the entire data objects to be created are known.
On the other hand, when the data objects are created one at a time and the total number of data objects is unknown, using hash tables is considered a good design choice.
Thus hash tables are used in PnetCDF to check conflicts of data objects defined within each process.
Using it to check conflicts across processes may not be the best choice, as shown in our results.
With all data objects to be created are known at the end-define stage, the sorting-based approach can outperform the hash table approach by a significant margin.


\subsection{New Header Format Approach Performance}

\label{sec:new-format-create}


\begin{figure}[t]
  \centering
  \includegraphics[width=.9\linewidth]{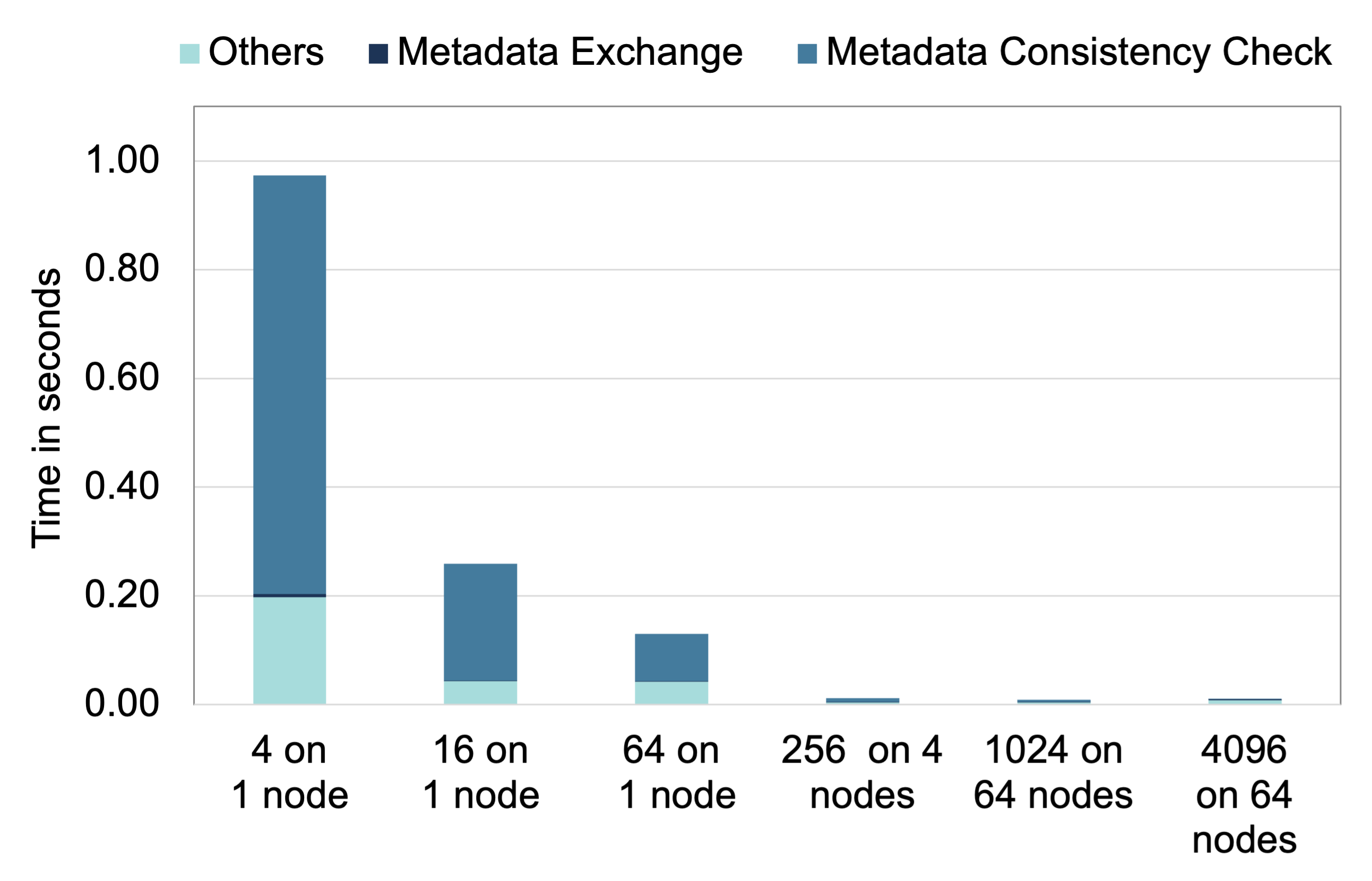}
  \caption{Timing breakdowns of the new header format approach using $dataset\_98M$.}
\label{fig:para-breakdown1}
\end{figure}

\begin{figure}[t]
  \centering
  \includegraphics[width=.9\linewidth]{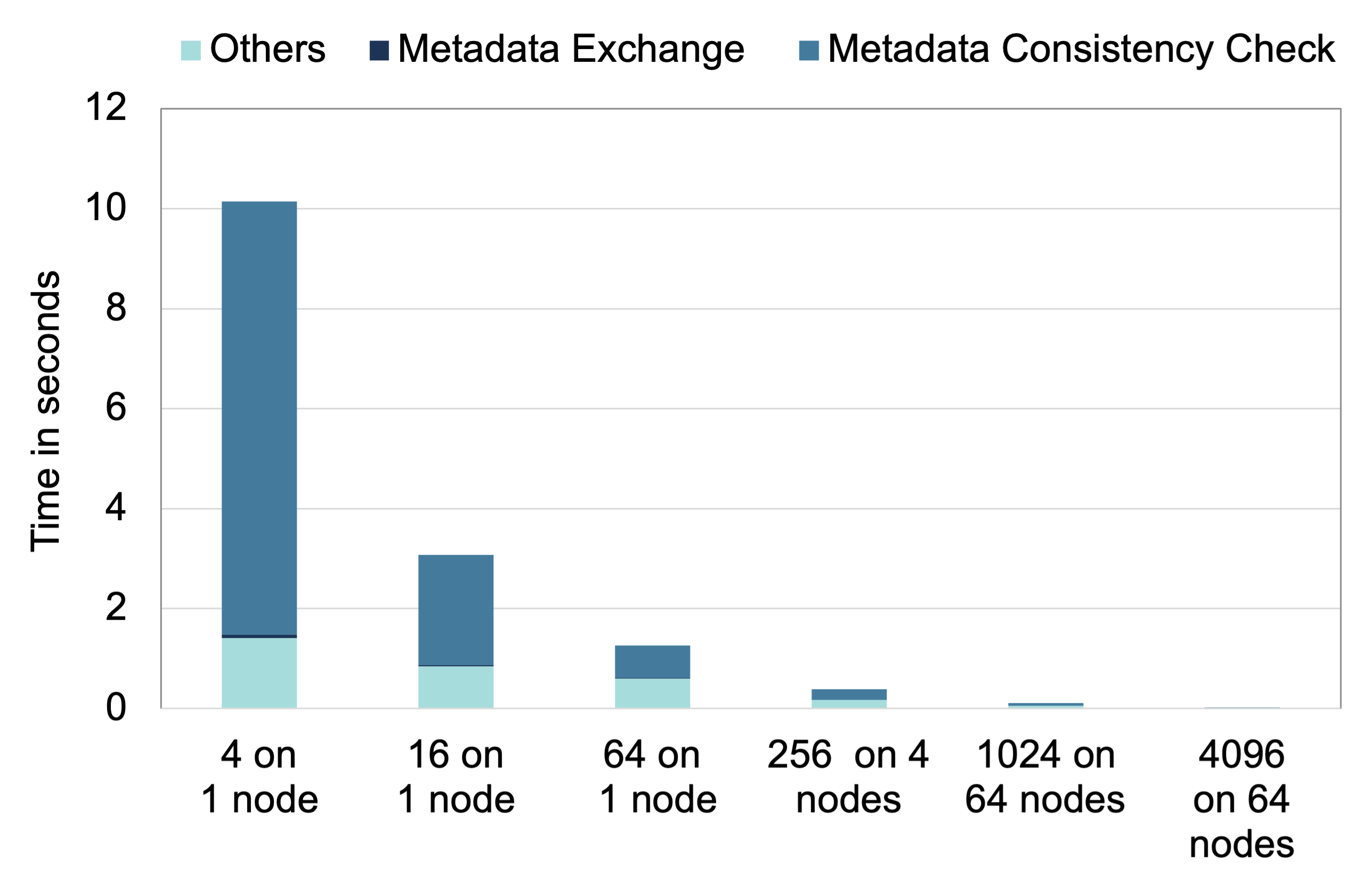}
  \caption{Timing breakdowns of the new header format approach using $dataset\_1G$.}
\label{fig:para-breakdown2}
\end{figure}

While Figures \ref{fig:write-time1} and \ref{fig:write-time2} show the total time of the new file header format approach (the rightmost bar in each group), its timing breakdowns are presented in Figures \ref{fig:para-breakdown1} and \ref{fig:para-breakdown2}, respectively. 
The total times show a significant performance improvements for the new file header format approach over other approaches. 
The timing breakdowns also reveal that the MPI communication becomes the dominant operation, unlike the three baseline approaches.
This cost proportionally decreases as the number application processes increases and the number of non-shared data objects per process decreases.

\begin{figure}[t]
  \centering
  \includegraphics[width=.9\linewidth]{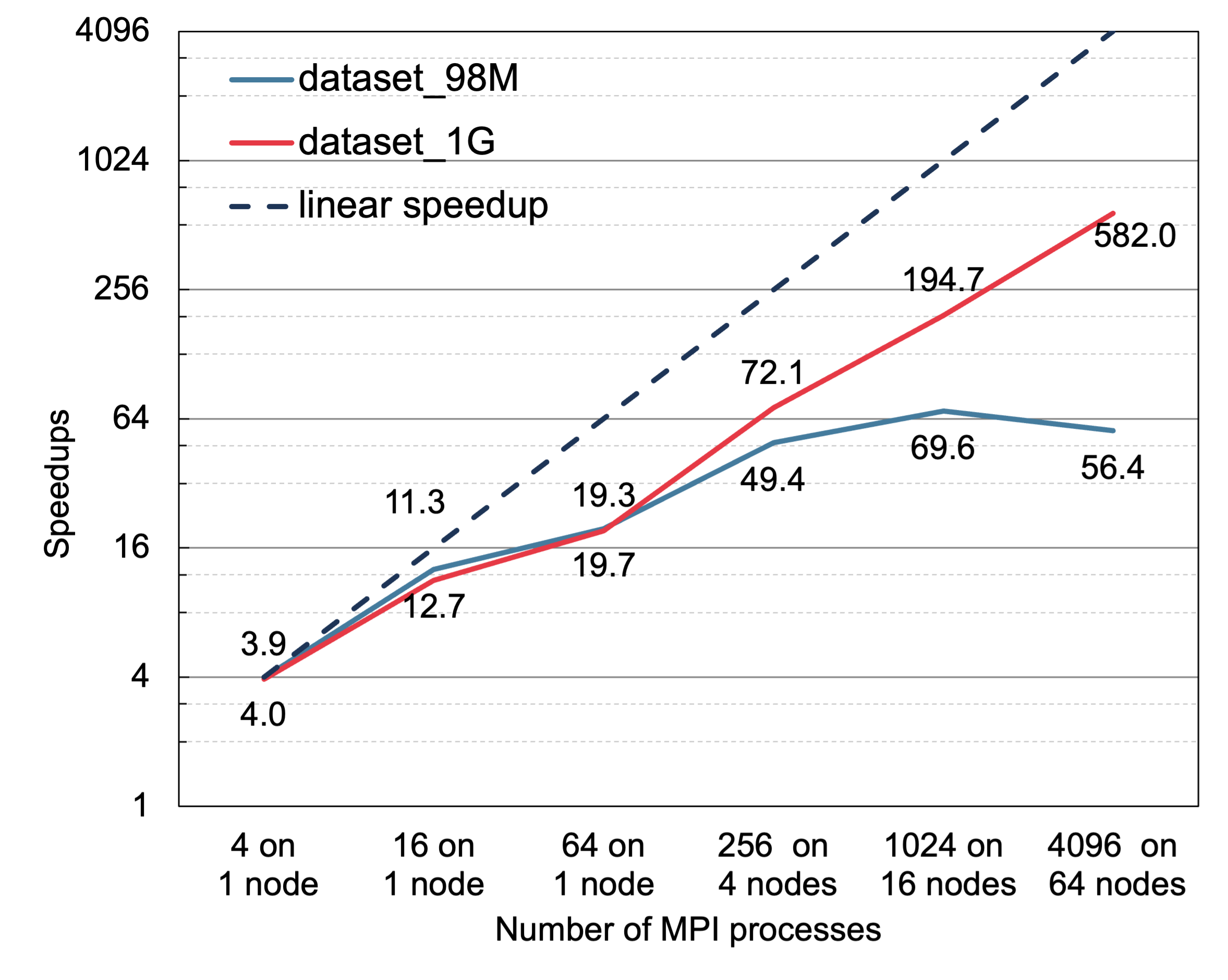}
  \caption{Speedups of the new header format approach. Speedups for other approaches are not shown because they are not scalable at all, as observed in Figure \protect{\ref{fig:write-time1}} \& \protect{\ref{fig:write-time2}}.}
\label{fig:speedup}
\end{figure}

We also observe that this approach achieves a better scalability in Figure \ref{fig:speedup} for both data sets. 
For the large data set $dataset\_1G$, a speedup of 582 is observed when running 4096 processes on 64 compute nodes.
For $dataset\_98M$, runtime performance also improves proportionally at smaller scales, but begins to plateau beyond the 256-process configuration.

As seen in Figures \ref{fig:para-breakdown1} and \ref{fig:para-breakdown2}, metadata consistency check remains the primary bottleneck at smaller scales. 
However, the cost decreases inversely proportional to the number of processes and the number of data objects per process decreases.
As described in Section \ref{sec:new-format}, data objects are jointly identified by their metadata blocks and their local identifiers within each metadata block. 
Thus, the runtime breakdowns align well with our expectations. 
The costs labeled as "others" include both file close and the metadata I/O.
They also exhibit reductions inversely proportional to the number of processes.
The former reduction is due to the fact that only metadata of non-shared objects allocated in each process's memory space need to be freed.
With less number of non-shared data objects decreases, there are less workload for cleanup and heap memory deallocation. 
For the latter, the reduction in the metadata I/O stage is attributed to the inherent parallelism in writing the metadata blocks. 
By enabling all processes to write the metadata to the file header, its performance becomes much better than having only the root process to write to the file.


Although metadata read operations are not the focus of this study, we would like to study the cost of reading the entire header from files in the proposed new format. 
We designed a weak-scaling experimental setting where each process reads the complete set of metadata into memory by iterating through all data objects.
Note that the applications are not required to read the entire metadata stored in a file.
Metadata of a data object can be read from the file only when it is demanded by an application process.
In the current PnetCDF library, the entire header is always read into a process's memory at the file open time.
In our experiment, we used an input file that contains 512 metadata blocks with the same data objects from $dataset\_98M$ and $dataset\_1G$ spread evenly among blocks.

Table \ref{tab:read-timings} summarizes the read operation performance for the new header format. 
The new header format approach demonstrates that the read performance is on par with that of the classic format with some variations. 
Specifically, for moderate header size ($dataset\_98M$), the new header format has slightly worse performance than the classic format, potentially due to the additional overheads in iterating through all metadata blocks. 
However, for $dataset\_1G$, the overheads of accessing metadata blocks become negligible and new header format exhibits a similar level of read performance. 

\begin{table}[t]
\small
\caption{Timings of reading the entire file header when using the proposed new header format.}
\label{tab:read-timings}
\resizebox{1.03\linewidth}{!}{%
\begin{tabular}{ccccc}
\hline
\multicolumn{5}{|c|}{\textit{dataset\_98M}} \\ \hline
\multicolumn{1}{|c|}{No. of processes}      & \multicolumn{1}{c|}{\begin{tabular}[c]{@{}c@{}}4 processes \\ on 1 node\end{tabular}} & \multicolumn{1}{c|}{\begin{tabular}[c]{@{}c@{}}64 processes\\ on 1 node\end{tabular}} & \multicolumn{1}{c|}{\begin{tabular}[c]{@{}c@{}}256 processes \\ on 4 nodes\end{tabular}}      & \multicolumn{1}{c|}{\begin{tabular}[c]{@{}c@{}}512 processes \\ on 8 nodes\end{tabular}} \\ \hline
\multicolumn{1}{|c|}{Classic netCDF format} & \multicolumn{1}{c|}{0.63}                                                     & \multicolumn{1}{c|}{1.06}                                                      & \multicolumn{1}{c|}{1.18}                                                           & \multicolumn{1}{c|}{1.28}                                                      \\ \hline
\multicolumn{1}{|c|}{New header format}     & \multicolumn{1}{c|}{1.00}                                                     & \multicolumn{1}{c|}{1.36}                                                      & \multicolumn{1}{c|}{1.47}                                                           & \multicolumn{1}{c|}{1.57}                                                      \\ \hline
\multicolumn{5}{c}{} \\ \hline
\multicolumn{5}{|c|}{\textit{dataset\_1G}} \\ \hline
\multicolumn{1}{|c|}{No. of processes}      & \multicolumn{1}{c|}{\begin{tabular}[c]{@{}c@{}}4 processes \\ on 1 node\end{tabular}} & \multicolumn{1}{c|}{\begin{tabular}[c]{@{}c@{}}64 processes\\ on 1 node\end{tabular}} & \multicolumn{1}{c|}{\begin{tabular}[c]{@{}c@{}}256 processes \\ on 4 nodes\end{tabular}} & \multicolumn{1}{c|}{\begin{tabular}[c]{@{}c@{}}512 processes \\ on 8 nodes\end{tabular}} \\ \hline
\multicolumn{1}{|c|}{Classic netCDF format} & \multicolumn{1}{c|}{6.31}                                                      & \multicolumn{1}{c|}{11.83}                                                      & \multicolumn{1}{c|}{12.34}                                                           & \multicolumn{1}{c|}{12.95}                                                      \\ \hline
\multicolumn{1}{|c|}{New header format}     & \multicolumn{1}{c|}{6.62}                                                      & \multicolumn{1}{c|}{10.04}                                                      & \multicolumn{1}{c|}{10.78}                                                           & \multicolumn{1}{c|}{10.66}                                                      \\ \hline
\multicolumn{1}{l}{}                        & \multicolumn{1}{l}{}                                                           & \multicolumn{1}{l}{}                                                            & \multicolumn{1}{l}{}                                                                 & \multicolumn{1}{l}{}                                                            \\
\multicolumn{1}{l}{}                        & \multicolumn{1}{l}{}                                                           & \multicolumn{1}{l}{}                                                            & \multicolumn{1}{l}{}                                                                 & \multicolumn{1}{l}{}                                                            \\
\multicolumn{1}{l}{}                        & \multicolumn{1}{l}{}                                                           & \multicolumn{1}{l}{}                                                            & \multicolumn{1}{l}{}                                                                 & \multicolumn{1}{l}{}                                                            \\
\multicolumn{1}{l}{}                        & \multicolumn{1}{l}{}                                                           & \multicolumn{1}{l}{}                                                            & \multicolumn{1}{l}{}                                                                 & \multicolumn{1}{l}{}                                                           
\end{tabular}%
}
\end{table}


\subsection{Memory Footprints}

\begin{figure*}[t]
  \centering
  \includegraphics[width=.85\linewidth]{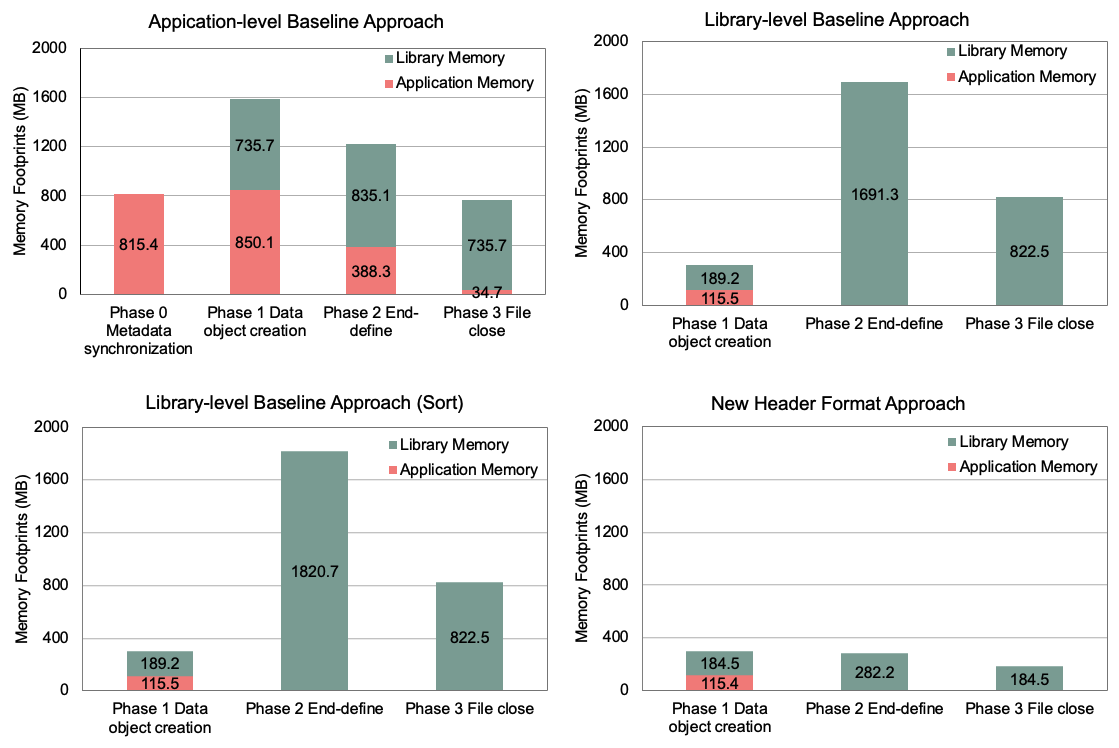}
  \caption{Memory footprints for all proposed approached using $dataset\_98M$ running 4 MPI processes on 1 node. The values are the sum of memory footprints of all processes.}
\label{fig:memory}
\end{figure*}

While runtime performance is a critical measure of the proposed approaches, memory footprint is equally important as it reflects the resource usage efficiency and scalability of each approach. 
Excessive memory usage can strain system resources, hinder scalability, and negatively impact the execution of other critical tasks in the applications. 
All previous results are based on up to 64 processes per node running on Perlmutter, as attempting to run 128 processes per node using $dataset\_1G$ caused all baseline approaches to fail due to the Out-of-Memory (OOM) error, whereas the new header format approach continued to run without OOM. 
To identify the processing steps that allocate excessive heap memory space, we collected the memory footprints of all the tests. 


Using $dataset\_98M$, we tracked down the memory usage when running 4 MPI processes on 1 node and show the results in Figure \ref{fig:memory}. 
From the bar chart of application-level baseline approach (top left), the memory allocated at the application level starts out high because every process must store a copy of metadata of all objects. 
Memory consumption reaches its high watermark during Phase 1 when metadata stored in the application is passed to the I/O library.
In the library-level baseline approach, the largest memory allocation occurs at Phase 2 during the synchronization of metadata. 
The overall high watermark of library-level baseline approach is about the same as the application-level approach.
There is an additional 7.6\% of memory usage for the library-level approach with the sorting optimization. 
This is due to the additional space required by constructing the data object name array for sorting step and checking for conflicts. 
In contrast, the new header format approach exhibits a significantly smaller memory footprint with only one-quarter of memory usage of the above baseline approaches.
The memory footprint only increases slightly in Phase 2, because processes only need to synchronize the index table and the metadata blocks containing shared data objects.
Without having to synchronize the metadata blocks containing non-shared data objects, the new format approach results in more efficient memory utilization.

In addition to memory footprints, we also discovered a scalability challenge in memory management caused by excessive small memory block allocations and fragmentation. 
The timings breakdowns in data object creation displays that the file close at Phase 3 becomes a non-negligible contributor to the end-to-end runtime. 
This issue arises from the fact that the number of memory free calls increases proportionally with the number of data objects and de-allocating them altogether at file close results in a noticeable cost.
A potential solution is to use a memory pool strategy, which pre-allocates a large memory chunk upfront to fulfill the small allocation requests. 
While this approach may improve performance, it may introduce some complexities, such as determining initial size, handling overflow and maintaining proper memory address alignment.

\section{Conclusion}

In this paper, we have demonstrated that data object creation with parallel I/O libraries can be expensive and become a performance bottleneck in large-scale dataset generation workflows where each process intends to write output arrays with descriptive information unique to other processes. 
One notable example is graph-structured dataset generation. Current parallel I/O libraries, such as HDF5 and PnetCDF, require metadata to be collectively created with identical metadata. 
This constraint becomes particularly problematic and runtime inefficient for parallel I/O libraries when creating data objects with different sizes among processes.

We progressively proposed to improve runtime efficiency and scalability in addressing this issue. 
The library-level baseline approach manages to remove the collective creation constraint in the I/O library and alleviate the applications from the burden of metadata synchronization. 
With the sort-based metadata consistency optimization, the performance of library-baseline approach can be further improved.
Among all the approaches studied in this paper, only the new header format approach achieves a scalable performance, as it allows each process to maintain only the non-shared data objects and thus be able to write metadata to the file header in parallel. 
These advancements pave the way for more efficient parallel I/O solutions, addressing metadata bottlenecks in scientific workflows with large-scale, heterogeneous datasets. 
As the data volumes produced and processed by applications in the HPC community continue to grow, ensuring parallel I/O libraries remain scalable and runtime efficient for large-scale, heterogeneous data has become increasingly important. 
Future works may include optimizing library memory management and the refinement of new header format.


\newpage

\bibliographystyle{ACM-Reference-Format}
\bibliography{citations}

\end{document}